\documentclass{sigchi-ext}
\usepackage[T1]{fontenc}
\usepackage{textcomp}
\usepackage[scaled=.92]{helvet} 
\usepackage{graphicx} 
\usepackage{balance} 
\usepackage{booktabs} 
\usepackage{ccicons} 
\usepackage{ragged2e} 
\usepackage{blindtext}

\def\plaintitle{SIGCHI Extended Abstracts Sample File: Note Initial
 Caps} 
\def\emptyauthor{}
\def\plainkeywords{Trust; Transparency; Recommender Systems; Mental Models; Explanation; Exploration}

\title{Trust and Transparency in Recommender Systems}

\numberofauthors{2}
\author{%
  \alignauthor{%
    \textbf{Clara Siepmann}\\
    \affaddr{Social Computing Group} \\
    \affaddr{Faculty of Engineering} \\
    \affaddr{University of Duisburg-Essen} \\
    \affaddr{Forsthausweg 2,  \\ 47057 Duisburg, Germany} \\
    \email{clara.siepmann@uni-due.de} } \alignauthor{%
    \textbf{Mohamed Amine Chatti}\\
    \affaddr{Social Computing Group} \\
    \affaddr{Faculty of Engineering} \\
    \affaddr{University of Duisburg-Essen} \\
    \affaddr{Forsthausweg 2,  \\ 47057 Duisburg, Germany} \\
    \email{mohamed.chatti@uni-due.de} } }

\definecolor{linkColor}{RGB}{6,125,233}
\hypersetup{%
 pdftitle={\plaintitle},
 pdfauthor={\emptyauthor},
 pdfkeywords={\plainkeywords},
 bookmarksnumbered,
 pdfstartview={FitH},
 colorlinks,
 citecolor=black,
 filecolor=black,
 linkcolor=black,
 urlcolor=linkColor,
 breaklinks=true,
}

\begin{document}


\maketitle
\RaggedRight{} 

\begin{abstract}
Trust is long recognized to be an important factor in Recommender Systems (RS). However, there are different perspectives on trust and different ways to evaluate it. Moreover, a link between trust and transparency is often assumed but not always further investigated. In this paper we first go through different understandings and measurements of trust in the AI and RS community, such as demonstrated and perceived trust. We then review the relationsships between trust and transparency, as well as mental models, and investigate different strategies to achieve transparency in RS such as explanation, exploration and exploranation (i.e., a combination of \textbf{explor}ation and expl\textbf{anation}). We identify a need for further studies to explore these concepts as well as the relationships between them. 

\end{abstract}





\section{Introduction}
Trust is no new concept in research regarding artificial intelligence (AI) \cite{hoffmanDynamicsTrustCyberdomains2009, lewandowskyDynamicsTrustComparing2000, ostromBehavioralTheoryLinking2003}. 
However trust is a multi-faceted concept shown by difference in definitions from disciplines such as psychology, sociology, and economics. In psychology, trust is defined as a cognitive attribute while sociology focus on trust as an attribute of human relations, and economics understand it as being calculated by humans \cite{kaur2022trustworthy}. 
This is further complicated by the fact that each application of AI has unique requirements regarding trust. For example, if we are looking at AI guided decision-making, it is important that the right amount of trust is induced by the system \cite{yang_how_2020, hemmer_effect_2022}. 
Trust also plays a role in Recommender Systems (RS) \cite{nunes2017systematic, he2016interactive}. In this context, trust and transparency are often linked, following the intuition that you will more likely trust a system that you can understand than one that is a black box to you. Transparency is often linked to users’ understanding of the RS inner logic and is supporting the user to build an accurate mental model of how the system works \cite{ngo2020exploring, kulesza2013too}. Moreover, providing transparency could enhance users’ trust in the system \cite{kunkel_let_2019, pu2012evaluating, zhao2019users}. However, some studies found that revealing
too much detail about the system’s inner logic may result in information overload, confusion, and a low level
of perceived understanding, which may in turn reduce users’ trust the system \cite{anannySeeingKnowingLimitations2018, hosseiniFourReferenceModels2018}. This suggests that there should be an optimal level of transparency which
will generate the highest level of users’ perceived trust in the system \cite{zhao2019users}. 
In this paper, we focus on trust and transparency in RS. We start by investigating how trust is conceptualized and measured in different systems as well as the role of transparency and mental models. Then, we address the relationship between trust and transparency, as well as related concepts including explanation and exploration. 
\section{Trust in AI}
\begin{marginfigure}[-25pc]
 \begin{minipage}{\marginparwidth}
  \centering
  \includegraphics[width=0.90\marginparwidth]{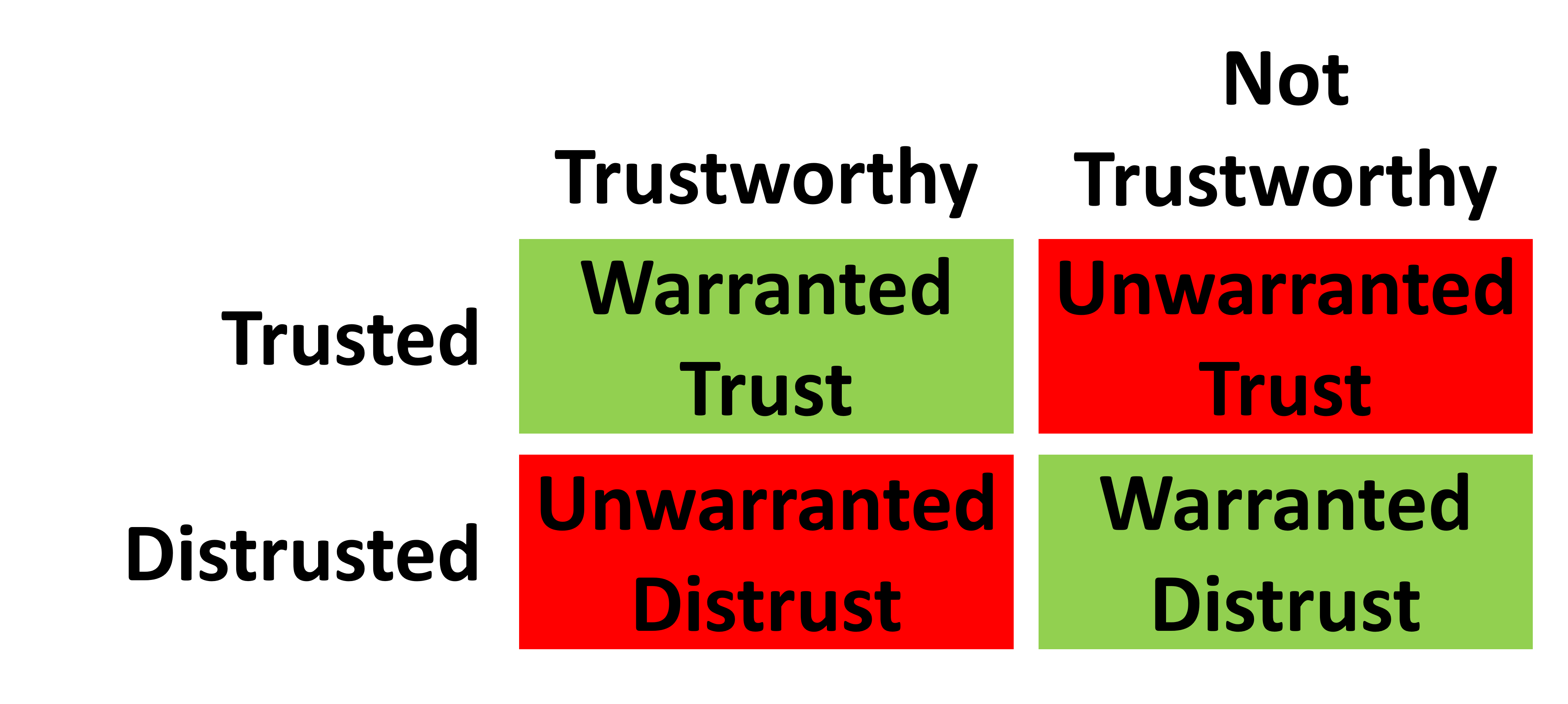}
  \caption{Possible kinds of trust in an AI systems as defined by Miller \protect\cite{miller_are_2022}}
  \label{fig:TrustMiller}
 \end{minipage}
\end{marginfigure}
\begin{marginfigure}[-7pc]
 \begin{minipage}{\marginparwidth}
  \centering
  \includegraphics[width=0.90\marginparwidth]{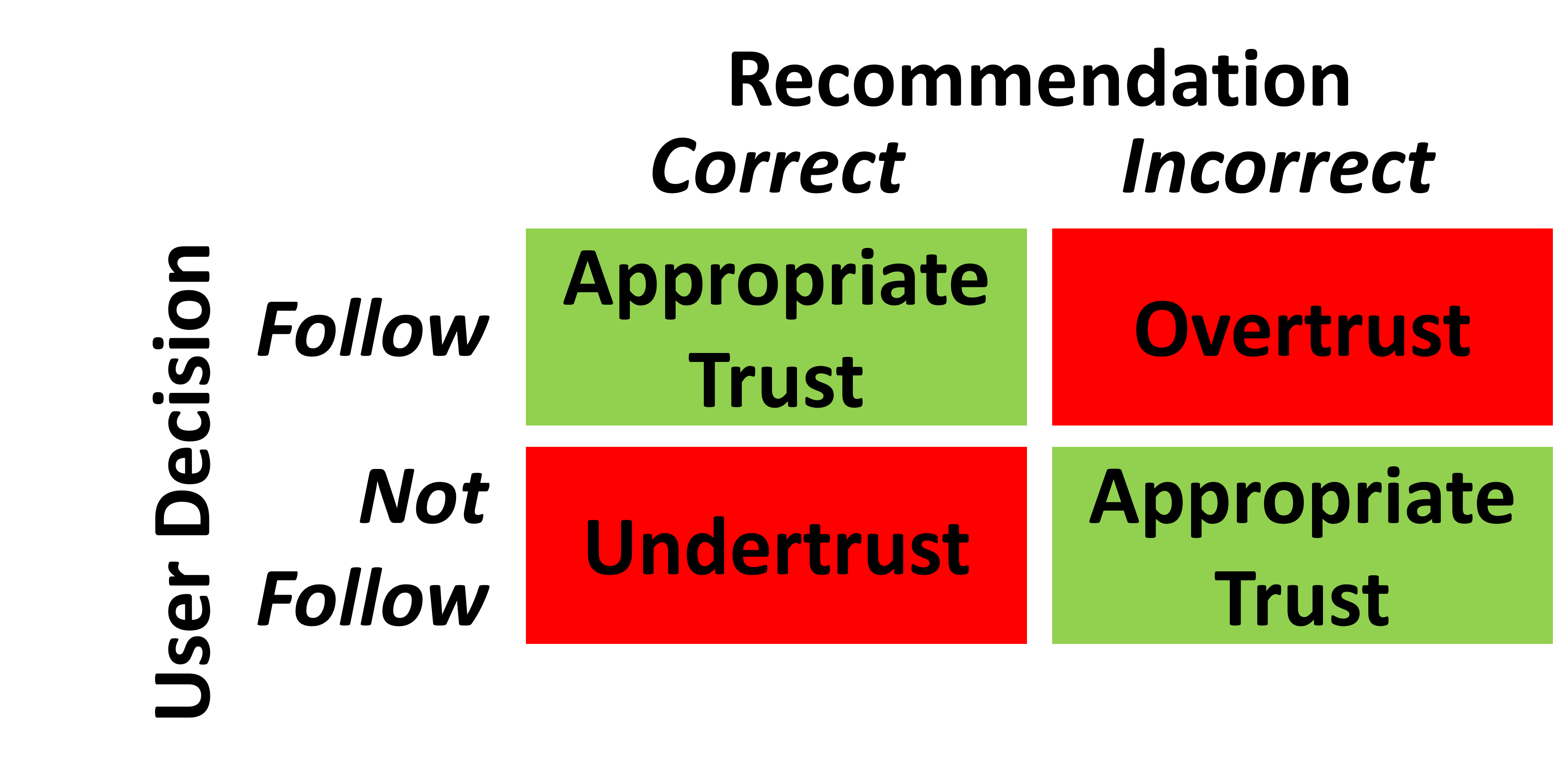}
  \caption{Possible kinds of trust in an AI systems as defined by Yang et al. - Figure taken from \protect\cite{yang_how_2020}}
  \label{fig:TrustYang}
 \end{minipage}
\end{marginfigure}

Miller \cite{miller_are_2022} differentiates between perceived trust and demonstrated trust. Perceived trust is measured through self-report of the user, while demonstrated trust is measured by observing the behaviour of the users. Questionnaires are often used to measure perceived trust. One popular questionnaire focusing on trust is the one by McKnight et al. \cite{mcknight_developing_2002}. Here trust is further split in four dimensions: \textit{Disposition to Trust}, \textit{Institution-Based Trust
}, \textit{Trusting Beliefs}, and \textit{Trusting Intentions} (see Figure \ref{fig:TrustMcKnight}). While the authors investigate trust in the context of e-commerce, this understanding of trust has been adapted in AI research \cite{Ghai2021explainable, vereschak2021evaluate}.

Demonstrated trust is especially important in the context of AI-guided decisions and collaborations between humans and AI. Miller \cite{miller_are_2022} defines four possible outcomes when we are inciting trust in a system depending on the trustworthiness of a system and the trust a user expresses: Warranted trust, unwarranted trust, unwarranted distrust, and warranted distrust. 
This follows the insight, that not all trust is good trust, especially if trust is followed by accepting a proposed decision of a system. The goal must be, that a user trusts trustworthy systems while distrusting not trustworthy systems (see Figure \ref{fig:TrustMiller}). The challenge now is to signal the trustworthiness of a system to the user. 
A very similar concept is used by Yang et al. \cite{yang_how_2020} (see Figure \ref{fig:TrustYang}). The authors apply it in the context of human-AI collaboration and use it to asses the quality of a decision made by a user with the help of an AI system. If a user correctly trusts/distrusts the recommended decision of a system, appropriate trust is achieved. 
\section{Trust in Recommender Systems}
Tintarev and Masthoff \cite{tintarev_survey_2007} identify trust as one of seven aims of explanations in RS. The authors conceptualize trust as "Increase users' confidence in the system" and propose two main approaches to measure trust, namely using questionnaires and measuring wanted effects such as customer loyalty or sales. This distinction can be mapped to the two kinds of trust proposed by Miller \cite{miller_are_2022}, where measuring effects is the equivalent to demonstrated trust and questionnaires measure perceived trust. 

\begin{marginfigure}[-7pc]
 \begin{minipage}{\marginparwidth}
  \centering
  \includegraphics[width=0.90\marginparwidth]{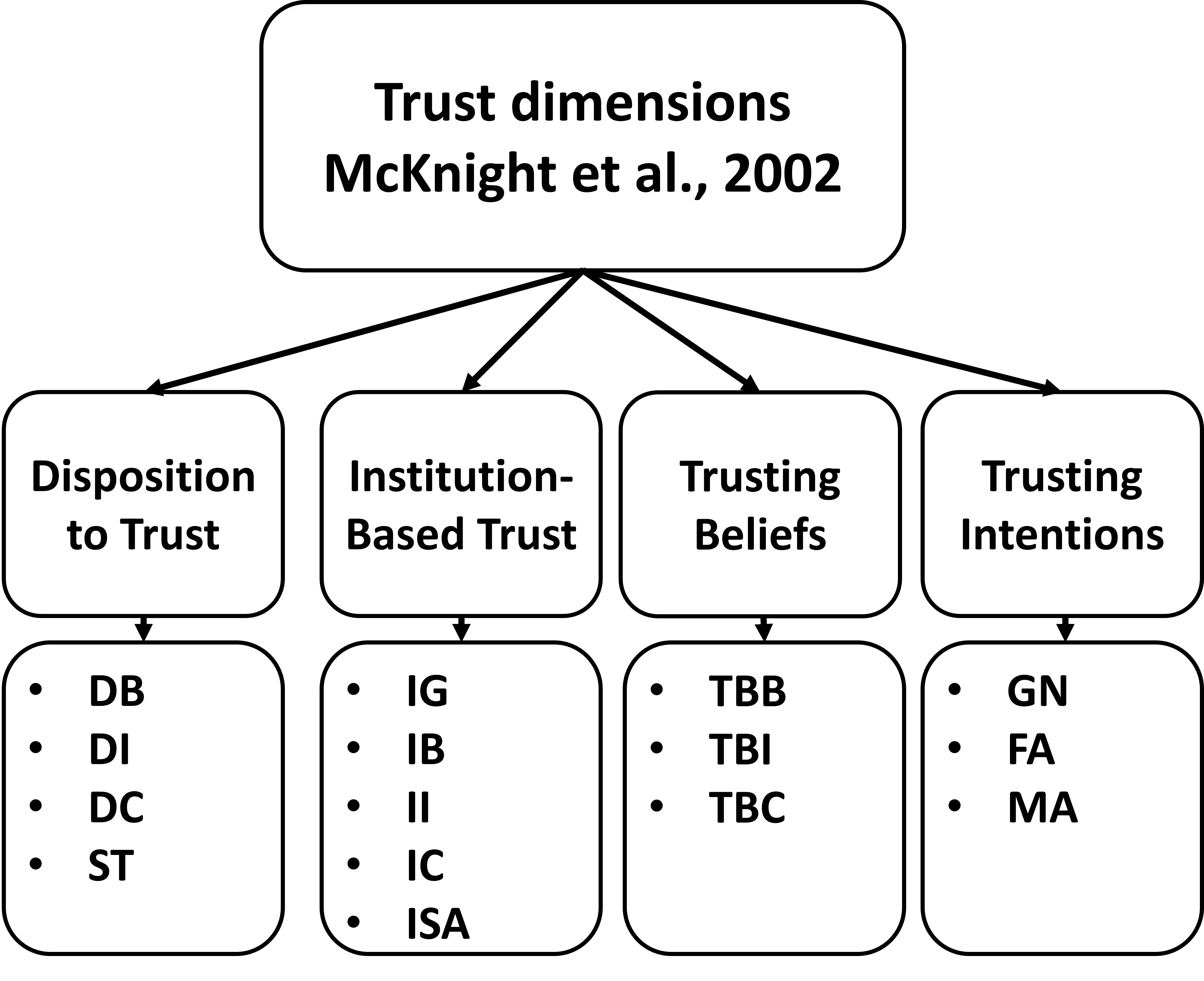}
  \caption{McKnight et al. \protect \cite{mcknight_developing_2002} divides trust in four dimensions. These are further split in \textit{Disposition to Trust}: Benevolence (DB), Integrity (DI), Competence (DC), Trusting Stance (ST); \textit{Institution-Based Trust}: Situational Normality-General (IG), Situational Normality-Benevolence (IB), Situational Normality-Integrity (II), Situational Normality-Competence (IC), Structural Assurance (ISA); \textit{Trusting Beliefs}: Benevolence (TBB), Integrity (TBI), Competence (TBC); \textit{Trusting Intentions}: Follow Advice (FA), Give Information (GI), Make Purchase (MA)}
  \label{fig:TrustMcKnight}
 \end{minipage}
\end{marginfigure}

One popular questionnaire in RS research is the ResQue framework introduced in \cite{pu_user-centric_2011} as a user-centric framework to measure different constructs such as quality, control/transparency, and attitudes in RS. In ResQue, trust is part of attitudes, i.e., user’s overall feeling towards an RS, and evaluated by a single item "The recommender can be trusted".
Another important framework is the one provided by Knijnenburg et al. \cite{knijnenburg_explaining_2012} who differentiate between \textit{Objective System Aspects (OSA)}, \textit{Subjective System Aspects (SSA)}, \textit{Situational Characteristics (SC)}, \textit{Experience (EXP)}, \textit{Interaction (INT)}, and \textit{Personal Characteristics (PC)}. According to the authors, trust can either be conceptualized as a PC, the general tendency to trust others, or as a SC, the trust in the RS. In the accompanying questionnaire the construct directly related to trust is the general trust in technology.
McKnight et al.'s concept is also adopted in the RS community \cite{hellmann_development_2022, kunkel_let_2019}. Each of the four dimensions is further split in sub-dimensions and each of these are evaluated using multiple items (see Figure \ref{fig:TrustMcKnight}).

In summary, there are multiple ways to measure trust in RS. But the concepts introduced by Miller \cite{miller_are_2022} and Yang et al. \cite{yang_how_2020}, such as warranted trust/ appropriate trust are to our knowledge not introduced in the evaluation of RS. 

\section{Transparency in Recommender Systems}
The connection between trust and transparency is drawn in the literature. For example Tintarev and Masthoff \cite{tintarev_survey_2007} argue that an increase in transparency should lead to an increase in trust. 
In their review, Nunes and Jannach \cite{nunes2017systematic} identified transparency as one of the key factors for users to develop trust. While this link is often assumed, the authors also identified a lack of explicit investigation of this relation.
Furthermore, transparency can be defined either from the system's perspective (\textit{objective transparency}) or from the user's perspective (\textit{subjective transparency}) \cite{zhao2019users}. While objective transparency centers the amount of information released by the system, subjective transparency means the awareness of the user that information regarding the working of the system is available. Additionally, the authors introduce \textit{user-perceived transparency}, which denotes the extent of which the users feel that they understood the provided information. 
It is also important to find balance in the amount of information given to the user as too much information can lead to a cognitive overload and negatively impact the perceived transparency \cite{zhao2019users}.

The importance of measuring the relationship between trust and transparency can be seen by the mixed results regarding the influence of transparency on trust. 
For example, Cramer et al. \cite{cramer_effects_2008} showed that transparency in a contend-based RS does not lead to an increase in trust. 
The link between transparency and trust can also be seen in \cite{hellmann_development_2022} where the authors included the effects of transparency on trust in their development of a questionnaire to measure the perceived transparency in a RS. They found a positive link between transparency and trust related measures.
Furthermore, we can distinguish between two main approaches for obtaining transparency as key to improve the trustworthiness of RS: \textbf{Transparency through explanation} and \textbf{transparency through exploration} \cite{shneiderman2022human}. 
\section{Mental Models in Recommender Systems}
Both explanation and exploration lead to the construction of mental models by users and therefore to transparency. A mental model describes the internal representation of an external system \cite{normanObservationsMentalModels1983}. This internal representation is not necessarily correct or complete and focuses on the practical implications rather than on technical details. This model is then used by the user to make assumptions about the behavior of a system as well as to interact with the system while pursuing a certain goal. Hence, an incorrect mental model leads to difficulties in the interaction with the system \cite{norman2013design}.
Given that trust is characterized by depending on another actor \cite{millerBehavioralMeasurementTrust2016}, it makes sense to assume that correct mental models lead to (appropriate) trust. The reason for this is that a correct mental model enables a user to correctly predict the behavior of a system and therefore enables the user to trust a system in the future \cite{ngo2020exploring}. 
\section{The Role of Explanation}
Opening the black-box of RS to end-users by providing explanations for system-generated recommendations has the potential to make RS transparent by helping users build an accurate mental model of how the RS works. Generally, explanations seek to show how a recommended item relates to user’s preferences. Explainable recommendation research covers a wide range of techniques and algorithms, and can be realized in many different ways \cite{nunes2017systematic}.
Explanation and transparency are often used interchangeably, as the assumption is, that explanation will lead to transparency \cite{tintarev2015explaining}. 
Given the close links between explanation and transparency on the one hand and transparency and trust on the other hand, the role of explanation is also investigated in relation to trust.
For instance, Kunkel et al. \cite{kunkel_let_2019} show that the quality of explanations has a positive effect on the trusting beliefs of the participants.
On the other hand, Chatti et al. \cite{chatti2022more} found that the perception of trust in an explainable RS is affected to different degrees by the explanation goal and user type.
\section{The Role of Exploration}
Recently, Shneiderman \cite{shneiderman2022human} suggests a shift from dependence on retrospective explanations to prospective user interfaces that are interactive, visual, and exploratory with the new goal to give users a better understanding of how the system works, so that they can prevent confusion and surprise that lead to the need of explanations. 
Exploratory user interfaces have the potential to guide users incrementally toward their goals and increase user control and transparency of AI processes. This can lead to building a more accurate mental model. For instance, visual analytics tools integrate interactive visual interfaces, the analytic capabilities of the computer, and the abilities of the user to allow novel discoveries and empower individuals to take control of the analytical
process; thus making the data analytics process transparent \cite{keim2008visual}. Moreover, various studies showed that human control and interaction can also contribute to increased transparency of AI/machine learning systems \cite{amershi2014power}. Furthermore, the benefits of exploratory user interfaces are well studied in the literature on interactive and more recently conversational RS \cite{he2016interactive,jugovac2017interacting,harambam2019designing,jannach2021survey}.      
In general, exploration can contribute to increased transparency and trust of the decision-making system as empowering users to take control of the system would help them build an accurate mental model of how the system works. Researchers recognized the potential of exploratory user interfaces to improve transparency \cite{tsai2017providing} and trust \cite{he2016interactive,jugovac2017interacting,harambam2019designing} in RS. However, while the relationship between explanation and trust is well investigated, only few studies explored the effects of exploration on trust. For example, the study in \cite{bakalov_approach_2013} showed that giving users control over their profile is not enough to increase trust. 

\section{The Combination of Explanation and Exploration}
Having explanation and exploration at the same time in the same RS is under-explored.
For instance, Tsai et al. \cite{tsai2021effects} investigated the way controllability and explanation in a user interface affect users' perception of trust in the RS. Even though trust was not the main focus of the study, the authors found a positive effect of controllability, explanation, as well as providing both of them at the same time on perceived trust. However, explanation and controllability were implemented as separate features in the RS. Another approach that could lead to interesting insights is the combination of both explanation and exploration as one feature in the RS leading to interactive explanations. 
\begin{marginfigure}[5pc]
 \begin{minipage}{\marginparwidth}
  \centering
  \includegraphics[width=1.1\columnwidth]{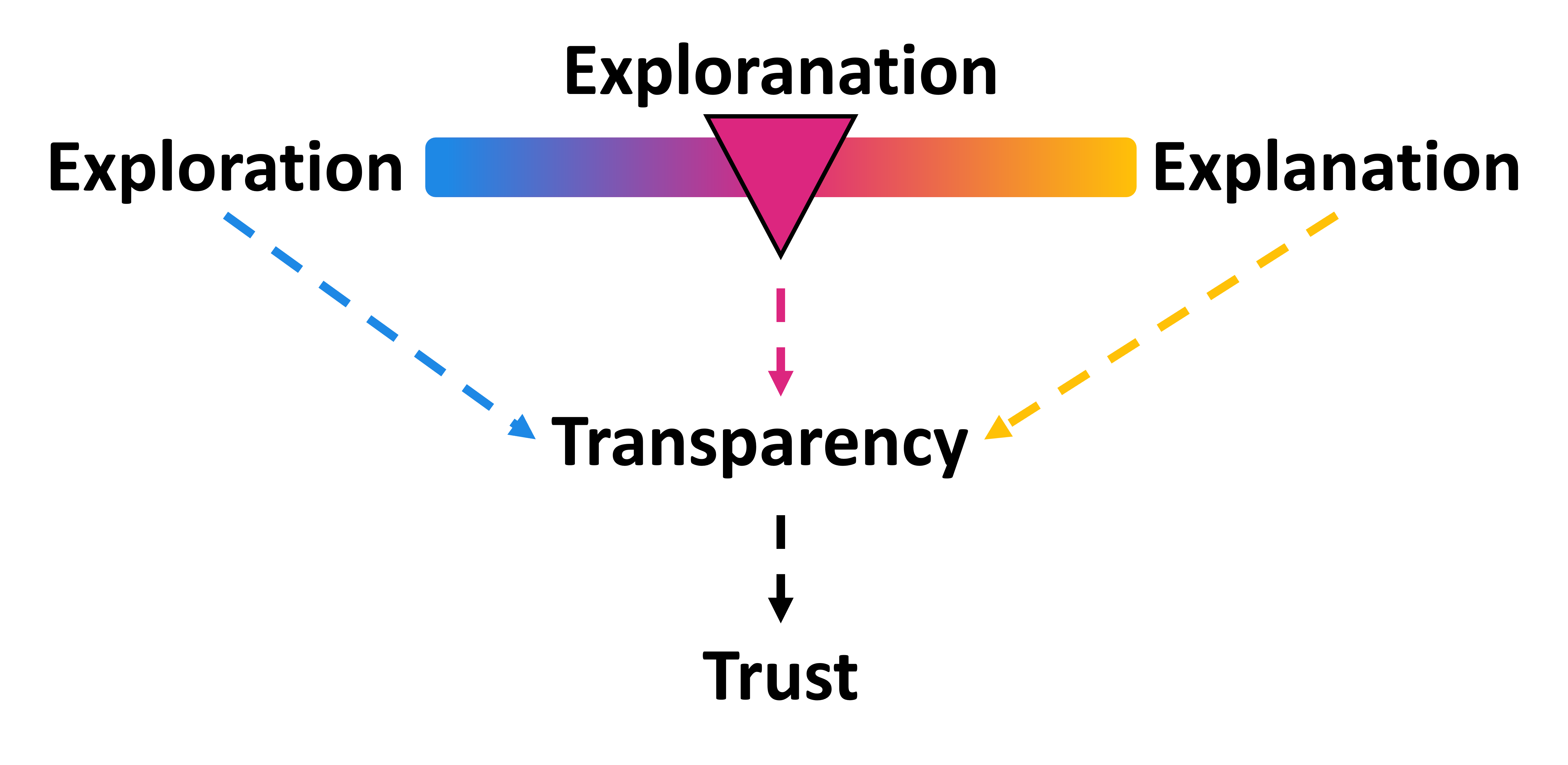}
    \caption{Exploration and Explanation are depicted as two ends of a continuum with Exploranations falling in the spectrum between them. The links between the concepts must be further investigated.}
    \label{fig:TypesRS}
 \end{minipage}
\end{marginfigure}
We borrow the term \textbf{exploranation} (i.e., a combination of \textbf{explor}ation and expl\textbf{anation}) introduced in \cite{ynnerman2018exploranation} to denote the way in which exploration can be effectively used to enrich explanation in RS. While explanation and exploration can be seen as two ends of a continuum, exploranation represents one point on the line from exploration to explanation, as depicted in Figure \ref{fig:TypesRS}. A similiar term for exploranation is interactive explanation \cite{guesmi2022if}. 
With the goal of bridging the gap between explainable AI (XAI) and human-computer interaction (HCI), research on designing and studying user interactions with XAI and explanatory machine learning has
emerged over the past few years \cite{cheng2019explaining, krause2016interacting, kuleszaPrinciplesExplanatoryDebugging2015a, sokol2020one}. Here the term interactive explanation is used to denote machine learning models, which enable the user to correct predictions of the system to mitigate learned biases and errors \cite{kuleszaPrinciplesExplanatoryDebugging2015a} as well as systems, that incorporate interactions in the explanatory machine learning pipeline to achieve transparency and trust \cite{tesoExplanatoryInteractiveMachine2019, schramowskiMakingDeepNeural2020}. 
Compared to the literature on XAI, little is known about how interactive explanation should be
designed and implemented in RS, so that explanation goals such as scrutability, transparency, trust, and user satisfaction
are met \cite{jannach2019explanations, hernandez2021effects}. Both in the literature and in real-world systems, there are only a few examples of RS that provide exploranation (i.e., interactive explanation), mainly to allow users to scrutinize the provided recommendations and correct the system’s assumptions \cite{balog2019transparent, guesmi2021open}, have a conversation, i.e., an exchange of questions and answers between the user and the system, using GUI-navigation or natural language conversation \cite{hernandez2021effects}, or interact with the explainable RS to receive \textit{What if} explanations \cite{guesmi2022if}. A distinction is to be made here between interactive recommendation and interactive explanation. While both empower users to take control of the RS process, they differ in the goal
of the control action. While the primary goal of interactive recommendation is to improve
and personalize the recommendation results, the goal of interactive explanation is to help both
the users for better understanding and the system designers for better model debugging. 
Given that both explanation and exploration can lead to transparency and trust, there is need to explore the relationships between these constructs (see Figure \ref{fig:TypesRS}). Moreover, it must be investigated in what specific cases we can reach what kind of trust through exploration, explanation, or exploranation.


\section{Challenges and Conclusion}
This short overview shows the complicated role trust plays in Recommender Systems (RS). Based on different perspectives on trust, we identified gaps in the current research on trust in RS. More focus should be laid on a more in depth evaluation of trust considering different dimensions. Moreover, it is needed to discuss the role that appropriate trust/warranted trust and demonstrated trust can play in RS, as well as how these concepts can be evaluated.
Also, given that we call for an investigation on trust, we must address the possibility of using the results for manipulating users into trusting systems or decisions, that should not be trusted. It is from the utmost importance to aim for an appropriate amount of trust in a system. But to achieve this goal of explainable AI, further research is needed. Additionally, different kinds of transparency (i.e., objective, subjective, user-perceived transparency) should be considered. 
Moreover, we shed light on the relationship between trust and transparency, as well as mental models, and how transparency can be achieved through explanation, exploration, or exploranation. We further call for more studies to explore the relationships between these aspects and perceived transparency and trust in RS, with their different dimensions. 
  
\balance{} 

\bibliographystyle{SIGCHI-Reference-Format}
\bibliography{sample}

\end{document}